\documentclass[aip,graphicx]{revtex4-1}
\newcommand\Ray{\mbox{\textit{Ra}}}  
\newcommand\Nus{\mbox{\textit{Nu}}} 
\draft 
\usepackage{graphicx}%
\usepackage{epstopdf, epsfig, enumitem,amsmath}
\usepackage{color}
\usepackage{hyperref}
\usepackage[normalem]{ulem}

\begin{document}


\title{Dependence of boundary layer thickness on layer height in laminar and transient convective regimes} 



\author{Andrei Sukhanovskii}
\email{san@icmm.ru}
 \author{Anna Evgrafova}%
 
\affiliation{ 
Institute of Continuous Media Mechanics, Perm, Russia
}%


\date{\today}

\begin{abstract}
Dependence of boundary layer thickness on layer height in laminar and transient regimes is studied for convection from localized heat source with an open surface. The measurements of Nusselt number and characteristic frequency of thermal plume formation are done for a wide range of Rayleigh number, different Prandtl numbers and different aspect ratios. The obtained results prove that for the developed (but not turbulent) convective flow  the boundary layer thickness is almost independent of the layer height and it is mainly defined by applied temperature difference  and physical properties of the fluid.
\end{abstract}

\pacs{47.27.te, 47.27.ep, 44.25.+f}

\maketitle 

\section{Introduction}

Thermal convection is a main source of the motion in the atmosphere and multiple technological systems. Convective heat transfer is more efficient than heat transfer provided by thermal conduction so understanding of different aspects of convective heat transfer (e.g. control and optimization) is of great importance. The classic and widely studied type of convection is Rayleigh-B\'enard  convection when the case of the cooled top and the heated bottom is considered \cite{goldstein1990high, siggia1994high, ahlers2009,chilla2012new,VermabookPBF}. Different aspects such as possibility of active control or a role of aspect ratio are studied \cite{howle1997active,wagner2013aspect}. Formation of large-scale geophysical flows provide interest to the horizontal convection when the sources of heating and cooling are located at the same height \cite{Mullarney2004,hughes2008horizontal, shishkina2016heat}. The need of enhancement of efficiency of coolers for technological systems of different scale such as electronic devices and nuclear reactors, optimization of fire safety requires understanding of specific features of heat transfer in the case of strongly non-homogeneous heating \cite{Torrance1969,1979Torrance,miroshnichenko2018turbulent, vasiliev2016high} or roughness \cite{toppaladoddi2017roughness,jiang2019convective,dong2020influence}.

The heat transport in convective systems is characterized by Nusselt number (the ratio of total heat flux with and without fluid motion). Numerous laboratory measurements \cite{goldstein1990high, siggia1994high, ahlers2009,chilla2012new} have clearly shown that the dependence of Nusselt number $\Nus$ on Rayleigh number $\Ray$ for Rayleigh-B\'enard  convection is defined by power laws. Understanding of theoretical grounds of these power laws is needed for correct parametrization of the convective heat transfer in large-scale industrial and geophysical flows. The key problem of convective heat transfer is understanding and description of thermal boundary layers \cite{siggia1994high}. Distinct thermal boundary layers in turbulent regimes provide most of the temperature drop throughout the whole layer, so the Nusselt number is proportional to the ratio of layer height to the thermal boundary layer thickness. Hence the thinner is thermal boundary layer the higher is the heat flux. There is a number of scaling theories describing dependence $\Nus(\Ray)$ as $\Nus\sim \Ray^{\beta}$ and providing different values of $\beta$ for different cases from $1/4$ at small $\Ray$ to $2/7$ or $1/3$ in developed turbulent state and achieving maximal value of $1/2$ in ultimate turbulence regime \cite{goldstein1990high, siggia1994high, ahlers2009, chilla2012new}. The horizontal convection is characterized by similar power laws but with different values of $\beta$ that varies from 1/6 to 1/3. The summary of different scalings for the horizontal convection is provided in \cite{shishkina2016heat} using the phase diagram in ($Ra, Pr$) coordinates. 

The early theory by Malkus for turbulent convection is based on the assumption that boundary layer thickness is satisfied to the critical boundary layer Rayleigh number, so it is independent on the layer height \cite{malkus1954heat}. It immediately leads to the scaling law $\Nus\sim \Ray^{1/3}$. Experimental studies usually satisfy to slightly less value of $\beta$. In a recent paper \cite{evgrafova2019specifics} it was stressed that scalings close to $2/7$  characterize the convective heat transfer for the quite different configurations of the heating and cooling distribution \cite{wright2017regimes,sukhanovsky2012horizontal,castaing1989scaling,chavanne2001turbulent}. 

In a present paper we want to check one important issue that was described in earlier studies but which is rarely taken into consideration. Early numerical studies (see references in \cite{golitsyn1979simple}) showed that even for small supercriticalities, when ratio $\Ray/\Ray_{critical}$ exceeds two, the pronounced thermal boundary layers are already formed. Also simple non-dimensional analysis in \cite{golitsyn1979simple} revealed that for large values of Nusselt number ($\Nus\gg1$), without making any assumption about the type of convection (laminar or turbulent), $\Nus$ depends on $\Ray$ as $\Nus\sim \Ray^{1/3}$ in the same way as it was proposed for the turbulent regimes in \cite{malkus1954heat}. Thus for relatively large $\Nus$ regardless of convection type the boundary layer thickness $\delta$ should be independent of the layer height. It is not clear is this scenario is valid in case of non-homogeneous heating, which is very important for multiple engineering applications. In order to check this assumption in case of localized heating, which is a specific case of mixed boundary conditions \cite{ripesi2014natural,wang2017thermal,bakhuis2018mixed}, we provide results for a wide range of Rayleigh number, different Prandtl numbers and different aspect ratios. 

\section{Experimental setup}
\label{sec:ExpSetup}

\begin{figure}
\includegraphics[width=.5\textwidth]{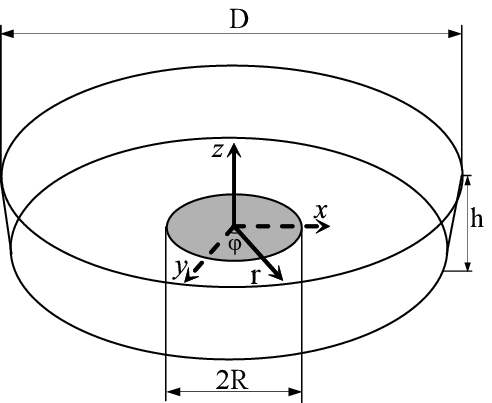}
\includegraphics[width=.9\textwidth]{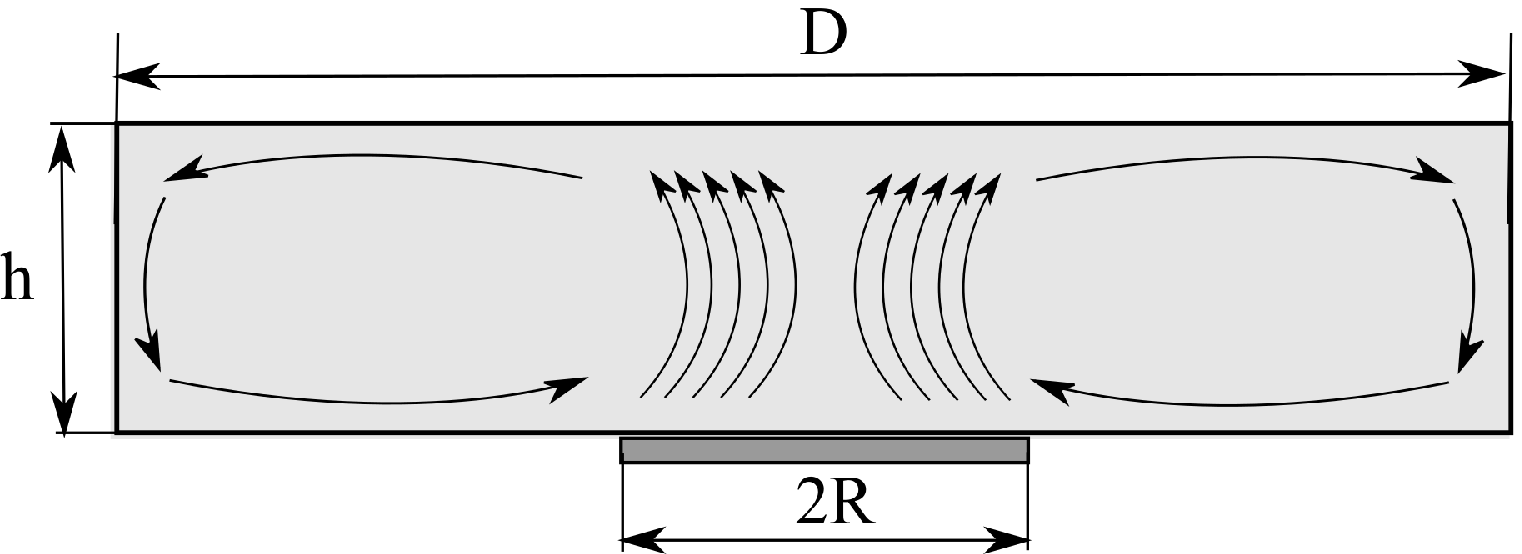}
\caption{Schemes of experimental setup (top) and large-scale circulation (bottom).}
\label{fig:expmodel}
\end{figure}

The detailed description of experimental setups and measurement systems can be found in \cite{evgrafova2019specifics} and here we provide only brief description. The general scheme of experimental setup is shown in Fig.\ref{fig:expmodel}(left). The experiments were carried out by cylindrical vessels of different diameter and size of the heater ($D=300$ mm, $2R=104$ mm -- model I, $D=690$ mm, $2R=100$ mm -- model II, $D=690$ mm, $2R=195$ mm -- model $\rm{II_b}$). The height of the fluid layer $h$ for model I was fixed at 30 mm and was varied for models II and $\rm{II_b}$  from 30 to 60 mm. The vessels were produced of Plexiglas. The heater is a brass or copper cylindrical plate mounted flush with the bottom and heated by an electrical coil placed on the lower side of the plate. The thermal condition on the heater was changed by varying the heat flux. For each experiment the heat flux was fixed and controlled. The temperature of the heater was measured by an built-in thermocouple. Silicon oils with different values of Prandtl number, PMS20, PMS10, PMS5 and PMS3 $(Pr = 209$, $Pr = 104$, $Pr = 67$ and $Pr=38$ at $T = 25^\circ $C) were used as working fluids. In all experiments the surface of the fluid was  open. The room temperature was kept constant during experiments by an air-conditioning system, and cooling of the fluid was provided mainly by the heat flux to the ambient air from the free surface. A horizontal row of twelve thermocouples was used for temperature measurements in the central area of the fluid. The distance between each thermocouple was 1 cm. For measurements at different heights the thermocouples were moved in vertical direction with a step of 1 mm by a motorized translation stage. For the estimation of the mean temperature of the fluid, one thermocouple was located at mid-height near the periphery.

The main non-dimensional parameters are Rayleigh number $\Ray$, Prandtl  number $\textit{Pr}$ and aspect ratio $a$.

\begin{equation}
 {{\Ray}}=\frac{g\alpha h^{3}\Delta T}{\nu k},  {{\textit{Pr}}} = \frac{\nu}k, {{a}} = \frac {h}{2R} \label{Ra}
\end{equation}

where $g$ is the gravitational acceleration, $\alpha$  is the coefficient of thermal expansion, $h$ is the layer height, $\nu$ is the coefficient of kinematic viscosity, and $k$ is the thermal diffusivity.
Convective heat transfer is described by the Nusselt number $Nu$.

\begin{equation}
 {{Nu}} = \frac{q}{q_c}\label{Nu}
\end{equation}

where $q_c$ is heat flux due to conduction, and $q$ is total heat flux. The measurements of the total heat flux were done directly. In Rayleigh-B\'enard  convection the boundary conditions are defined using a constant temperature difference or heat flux. In the present configuration, that is often considered for laboratory modeling of geophysical flows \cite{batalov2010laboratory,sukhanovskii2016,sukhanovskii2016laboratory,sukhanovskii2017non,sukhanovskii2020importance} , the measurements of Nusselt number are more complex due to open surface of the fluid. Because of that the temperature at the top boundary depends on ambient conditions and it is unclear how $q_c$ should be evaluated. Here we proposed to calculate the heat flux $q_c$ as follows:

\begin{equation}
 {q_c = \lambda\frac{T_h - T_s}{h}}\label{flux}
\end{equation}

where $\lambda$ is the thermal conductivity, $T_h$ is the temperature of the heater, and $T_s$ is the temperature of the fluid surface. The measurements of the Nusselt number were done for the quasi-stationary regime when the mean temperature of the fluid was assumed constant. It should be noted that due to the small-scale convection instantaneous local temperature measurements always reveal small pulsations.  The row of thermocouples \cite{evgrafova2019specifics} was used for measurements of the fluid surface temperature $T_s$ .

\section{Results}
\label{sec:Res}

Local heating in the closed tank always produces large-scale circulation. In the present configuration the heating in the bottom center results in the formation of an intensive buoyant flow. At the open surface the divergent warm fluid is cooled and goes downward, mostly near the side wall. Large-scale flow occupies the whole vessel. The structure of the main flow is shown in Fig.\ref{fig:expmodel}(right). The mean flow is convergent in the lower part of the vessel and divergent in the upper layer. The relatively cold advective flow forms a thermal boundary layer with a considerable negative (unstable) temperature gradient over the hot plate that results in appearance of thermal plumes and formation of horizontal rolls. Recent studies showed that formation and characteristics of small-scale flows strongly depend on the heating intensity \cite{batalov2007experimental,sukhanovsky2012horizontal,sukhanovskii2016}.

\begin{figure}
\center{\includegraphics[width=0.95\linewidth]{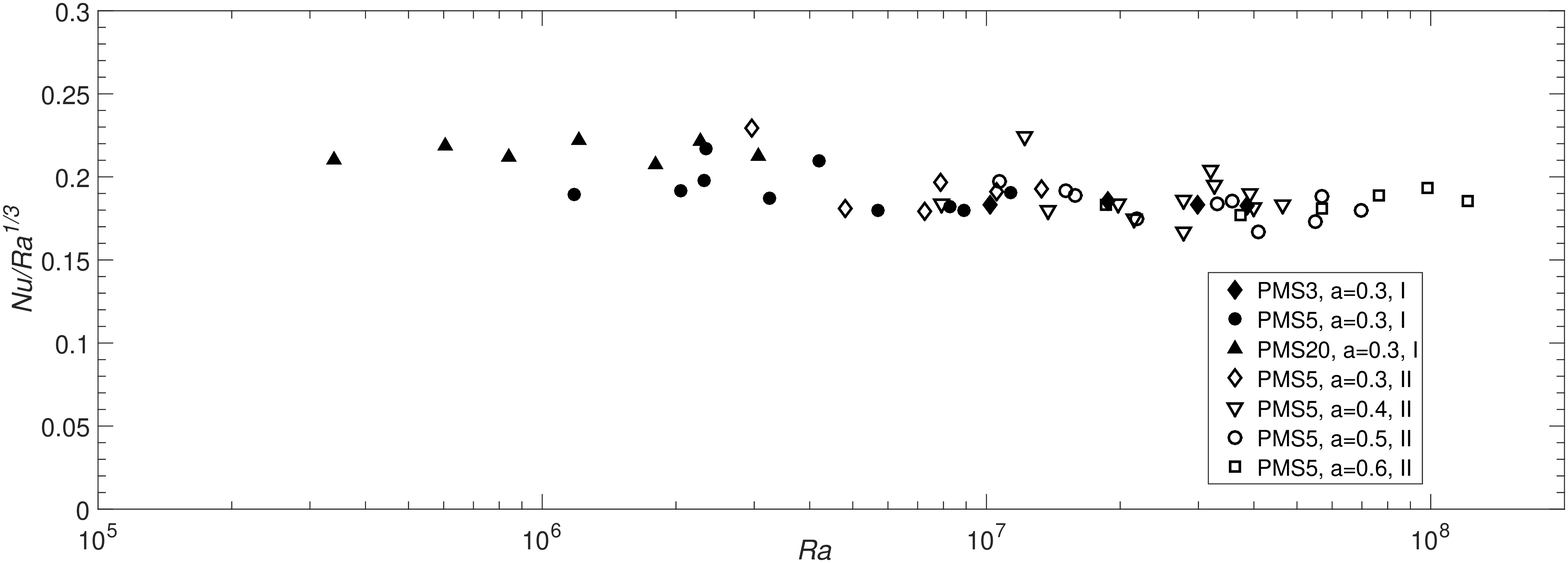}\\}
\caption{Dependence of reduced Nusselt number on Rayleigh number $\Ray$ for different fluids (from PMS3 to PMS20) and aspect ratios (from $a=0.3$ to $a=0.6$); I and II -- the type of the experimental model, solid line shows scaling $\sim \Ray^{1/3}$.}
\label{Nu_Ra}
\end{figure}

At first we consider dependence of reduced Nusselt number on Rayleigh number for different fluids and aspect ratios (Fig.\ref{Nu_Ra}). Reduced Nusselt number is very sensitive to the deviation from the chosen scaling and we see that our data is very close to $\Nus\sim \Ray^{1/3}$. The best fit for the whole set of data is $\Nus\sim \Ray^{0.31\pm0.015}$ which is remarkably close to the steady solutions \cite{waleffe2015heat}  up to $\Ray<10^9$ and 3D turbulent data \cite{niemela2000turbulent,he2012transition} up to $\Ray<10^{11}$. It is evident that variation of the physical properties of the fluid and aspect ratio $a$ does not lead to the substantial variation of the Nusselt number. Using $\Nus\sim h/\delta$ and $\Nus\sim \Ray^{1/3}$ we obtain $\delta\sim {\Delta T}^{-1/3}$, so for the considered case of developed but not turbulent convection the thermal layer thickness practically does not depend on the layer height and it is mostly defined by the applied temperature difference $\Delta T$ and physical properties of the fluid (kinematic viscosity and thermal diffusivity). 

\begin{figure} 
\center{\includegraphics[width=0.95\linewidth]{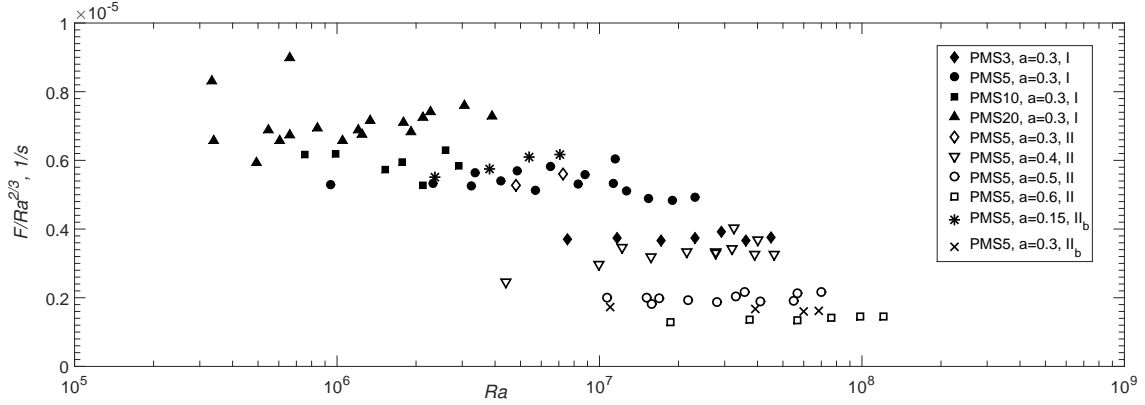}\\}
\caption{Dependence of reduced characteristic frequency $F$ on Rayleigh number $\Ray$, for different fluids (from PMS3 to PMS20) and aspect ratios (from $a=0.15$ to $a=0.6$); I,II and $\rm{II_b}$ -- the type of the experimental model.}
\label{F}
\end{figure}

\begin{figure}
\center{\includegraphics[width=0.95\linewidth]{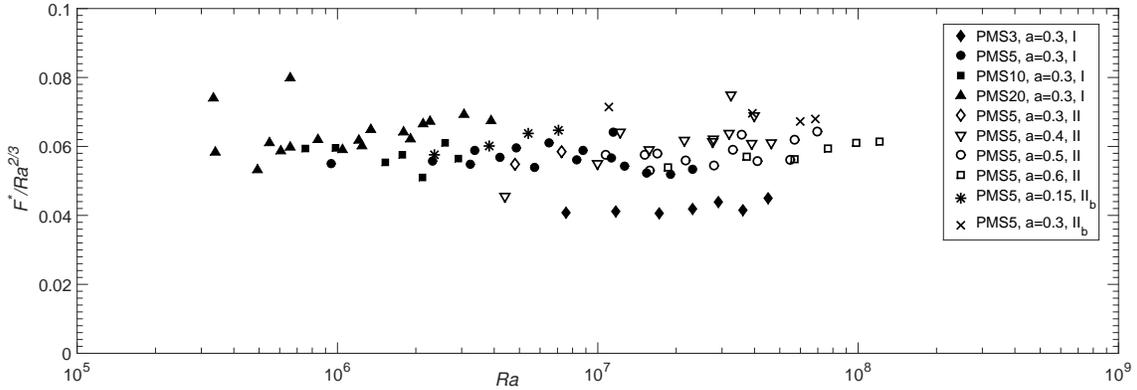}\\}
\caption{Dependence of non-dimensional reduced characteristic frequency $F\times h^2/k$ on Rayleigh number $\Ray$, for different fluids (from PMS3 to PMS20) and aspect ratios (from $a=0.15$ to $a=0.6$); I,II and $\rm{II_b}$ -- the type of the experimental model, solid line shows scaling $\sim \Ray^{2/3}$.}
\label{Fcorr}
\end{figure}

There is alternative way for the estimation of the thermal boundary layer thickness. For the considered range of parameters the process of thermal plume formation is periodic \cite{sukhanovskii2016,evgrafova2019specifics}. Following \cite{cioni1997strongly} the characteristic frequency of the plume formation can be estimated as $F \sim k/\delta^2$. Dependence of $F$ on Rayleigh number for different values of Prandtl number and aspect ratios is shown in Fig.\ref{F}. It is clear that the values of $F$ for different fluids are defined by similar scaling $F\sim \Ray^{\gamma}$ but prefactor definitely depends on properties of the fluid and layer height. Indeed if we consider non-dimensional frequency $F^*=F\times h^2/k$ then almost all points belong to the one curve with a scaling close to $\Ray^{2/3}$ (Fig.\ref{Fcorr}). The experimental points for the fluid with the lowest value of Prandtl number (PMS3) lie a bit lower, which means that for the low viscous fluids additional factor (value of Prandtl number) should be taken into account. Using $F^*\sim \Ray^{2/3}$ and $F\sim k/\delta^2$ we immediately obtain that boundary layer thickness depends on temperature difference ($\delta\sim {\Delta T}^{-1/3}$) but does not depend on the layer height.

\section{Conclusions}
\label{sec:conc}

The main goal of this study is checking the assumption that for relatively large values of the Nusselt number regardless of the convection type the boundary layer thickness is independent of the layer height. For this purpose we carry out measurements of Nusselt number and characteristic frequency of thermal plume formation for convection from localized heat source with an open surface. The measurements are done for a wide range of Rayleigh number, different Prandtl numbers and aspect ratios. 

The obtained results prove that for the developed (but not turbulent) convective flow, when $\Ray$ is substantially higher than $\Ray_{critical}$ the boundary layer thickness is almost independent of the layer height and it is mainly defined by applied temperature difference $\Delta T$ and physical properties of the fluid (kinematic viscosity and thermal diffusivity). It agrees with the results of \cite{batalov2007experimental} where independence of the boundary layer thickness of the layer height  was revealed for the different configuration (rectangular vessel with partially heated and cooled bottom). Our results are also in a good agreement with the fundamental result of \cite{golitsyn1979simple} that characteristic convective time does not depend on the layer height.

The loss of dependence of $\delta$ on $h$ results in scaling $\Nus\sim \Ray^{1/3}$ and $F^*\sim \Ray^{2/3}$ and this is probably the most simple explanation why the similar scalings were obtained for the quite different configurations of the heating and cooling distribution in cylindrical and rectangular vessels  \cite{wright2017regimes,evgrafova2019specifics,sukhanovsky2012horizontal,castaing1989scaling}. Also we can assume that described scenario is characteristic for a variety of convective systems when the size of the heating (cooling) area is sufficiently large in comparison with the the thickness of the thermal boundary layer even in the case of strongly non-homogeneous spatial distribution of the heating (cooling). This is very important for optimization of the heat transfer using mixed boundary conditions. The fact that scaling in experimental studies is slightly lower than 1/3 and is more close to 2/7 means that there is still a weak dependence of thermal boundary layer thickness on the layer height which is vanished in developed turbulent regimes.

\begin{acknowledgments}
The financial support of the government programme is gratefully acknowledged.

We thank for useful comments Detlef Lohse.
\end{acknowledgments}

\bibliography{san}

\begin{thebibliography}{37}%
\makeatletter
\providecommand \@ifxundefined [1]{%
 \@ifx{#1\undefined}
}%
\providecommand \@ifnum [1]{%
 \ifnum #1\expandafter \@firstoftwo
 \else \expandafter \@secondoftwo
 \fi
}%
\providecommand \@ifx [1]{%
 \ifx #1\expandafter \@firstoftwo
 \else \expandafter \@secondoftwo
 \fi
}%
\providecommand \natexlab [1]{#1}%
\providecommand \enquote  [1]{``#1''}%
\providecommand \bibnamefont  [1]{#1}%
\providecommand \bibfnamefont [1]{#1}%
\providecommand \citenamefont [1]{#1}%
\providecommand \href@noop [0]{\@secondoftwo}%
\providecommand \href [0]{\begingroup \@sanitize@url \@href}%
\providecommand \@href[1]{\@@startlink{#1}\@@href}%
\providecommand \@@href[1]{\endgroup#1\@@endlink}%
\providecommand \@sanitize@url [0]{\catcode `\\12\catcode `\$12\catcode
  `\&12\catcode `\#12\catcode `\^12\catcode `\_12\catcode `\%12\relax}%
\providecommand \@@startlink[1]{}%
\providecommand \@@endlink[0]{}%
\providecommand \url  [0]{\begingroup\@sanitize@url \@url }%
\providecommand \@url [1]{\endgroup\@href {#1}{\urlprefix }}%
\providecommand \urlprefix  [0]{URL }%
\providecommand \Eprint [0]{\href }%
\providecommand \doibase [0]{http://dx.doi.org/}%
\providecommand \selectlanguage [0]{\@gobble}%
\providecommand \bibinfo  [0]{\@secondoftwo}%
\providecommand \bibfield  [0]{\@secondoftwo}%
\providecommand \translation [1]{[#1]}%
\providecommand \BibitemOpen [0]{}%
\providecommand \bibitemStop [0]{}%
\providecommand \bibitemNoStop [0]{.\EOS\space}%
\providecommand \EOS [0]{\spacefactor3000\relax}%
\providecommand \BibitemShut  [1]{\csname bibitem#1\endcsname}%
\let\auto@bib@innerbib\@empty
\bibitem [{\citenamefont {Goldstein}, \citenamefont {Chiang},\ and\
  \citenamefont {See}(1990)}]{goldstein1990high}%
  \BibitemOpen
  \bibfield  {author} {\bibinfo {author} {\bibfnamefont {R.}~\bibnamefont
  {Goldstein}}, \bibinfo {author} {\bibfnamefont {H.}~\bibnamefont {Chiang}}, \
  and\ \bibinfo {author} {\bibfnamefont {D.}~\bibnamefont {See}},\ }\bibfield
  {title} {\enquote {\bibinfo {title} {High-rayleigh-number convection in a
  horizontal enclosure},}\ }\href@noop {} {\bibfield  {journal} {\bibinfo
  {journal} {Journal of Fluid Mechanics}\ }\textbf {\bibinfo {volume} {213}},\
  \bibinfo {pages} {111--126} (\bibinfo {year} {1990})}\BibitemShut {NoStop}%
\bibitem [{\citenamefont {Siggia}(1994)}]{siggia1994high}%
  \BibitemOpen
  \bibfield  {author} {\bibinfo {author} {\bibfnamefont {E.~D.}\ \bibnamefont
  {Siggia}},\ }\bibfield  {title} {\enquote {\bibinfo {title} {High rayleigh
  number convection},}\ }\href@noop {} {\bibfield  {journal} {\bibinfo
  {journal} {Annual review of fluid mechanics}\ }\textbf {\bibinfo {volume}
  {26}},\ \bibinfo {pages} {137--168} (\bibinfo {year} {1994})}\BibitemShut
  {NoStop}%
\bibitem [{\citenamefont {Ahlers}, \citenamefont {Grossmann},\ and\
  \citenamefont {Lohse}(2009)}]{ahlers2009}%
  \BibitemOpen
  \bibfield  {author} {\bibinfo {author} {\bibfnamefont {G.}~\bibnamefont
  {Ahlers}}, \bibinfo {author} {\bibfnamefont {S.}~\bibnamefont {Grossmann}}, \
  and\ \bibinfo {author} {\bibfnamefont {D.}~\bibnamefont {Lohse}},\ }\bibfield
   {title} {\enquote {\bibinfo {title} {Heat transfer and large scale dynamics
  in turbulent rayleigh-b{\'e}nard convection},}\ }\href@noop {} {\bibfield
  {journal} {\bibinfo  {journal} {Reviews of modern physics}\ }\textbf
  {\bibinfo {volume} {81}},\ \bibinfo {pages} {503} (\bibinfo {year}
  {2009})}\BibitemShut {NoStop}%
\bibitem [{\citenamefont {Chill{\`a}}\ and\ \citenamefont
  {Schumacher}(2012)}]{chilla2012new}%
  \BibitemOpen
  \bibfield  {author} {\bibinfo {author} {\bibfnamefont {F.}~\bibnamefont
  {Chill{\`a}}}\ and\ \bibinfo {author} {\bibfnamefont {J.}~\bibnamefont
  {Schumacher}},\ }\bibfield  {title} {\enquote {\bibinfo {title} {New
  perspectives in turbulent rayleigh-b{\'e}nard convection},}\ }\href@noop {}
  {\bibfield  {journal} {\bibinfo  {journal} {The European Physical Journal E}\
  }\textbf {\bibinfo {volume} {35}},\ \bibinfo {pages} {58} (\bibinfo {year}
  {2012})}\BibitemShut {NoStop}%
\bibitem [{\citenamefont {Verma}(2018)}]{VermabookPBF}%
  \BibitemOpen
  \bibfield  {author} {\bibinfo {author} {\bibfnamefont {M.~K.}\ \bibnamefont
  {Verma}},\ }\href {\doibase 10.1142/10928} {\emph {\bibinfo {title} {Physics
  of Buoyant Flows: From Instabilities to Turbulence}}}\ (\bibinfo  {publisher}
  {World Scientific},\ \bibinfo {address} {Singapore},\ \bibinfo {year}
  {2018})\BibitemShut {NoStop}%
\bibitem [{\citenamefont {Howle}(1997)}]{howle1997active}%
  \BibitemOpen
  \bibfield  {author} {\bibinfo {author} {\bibfnamefont {L.~E.}\ \bibnamefont
  {Howle}},\ }\bibfield  {title} {\enquote {\bibinfo {title} {Active control of
  rayleigh--b{\'e}nard convection},}\ }\href@noop {} {\bibfield  {journal}
  {\bibinfo  {journal} {Physics of Fluids}\ }\textbf {\bibinfo {volume} {9}},\
  \bibinfo {pages} {1861--1863} (\bibinfo {year} {1997})}\BibitemShut {NoStop}%
\bibitem [{\citenamefont {Wagner}\ and\ \citenamefont
  {Shishkina}(2013)}]{wagner2013aspect}%
  \BibitemOpen
  \bibfield  {author} {\bibinfo {author} {\bibfnamefont {S.}~\bibnamefont
  {Wagner}}\ and\ \bibinfo {author} {\bibfnamefont {O.}~\bibnamefont
  {Shishkina}},\ }\bibfield  {title} {\enquote {\bibinfo {title} {Aspect-ratio
  dependency of rayleigh-b{\'e}nard convection in box-shaped containers},}\
  }\href@noop {} {\bibfield  {journal} {\bibinfo  {journal} {Physics of
  Fluids}\ }\textbf {\bibinfo {volume} {25}},\ \bibinfo {pages} {085110}
  (\bibinfo {year} {2013})}\BibitemShut {NoStop}%
\bibitem [{\citenamefont {{Mullarney}}, \citenamefont {{Griffiths}},\ and\
  \citenamefont {{Hughes}}(2004)}]{Mullarney2004}%
  \BibitemOpen
  \bibfield  {author} {\bibinfo {author} {\bibfnamefont {J.~C.}\ \bibnamefont
  {{Mullarney}}}, \bibinfo {author} {\bibfnamefont {R.~W.}\ \bibnamefont
  {{Griffiths}}}, \ and\ \bibinfo {author} {\bibfnamefont {G.~O.}\ \bibnamefont
  {{Hughes}}},\ }\bibfield  {title} {\enquote {\bibinfo {title} {{Convection
  driven by differential heating at a horizontal boundary}},}\ }\href {\doibase
  10.1017/S0022112004000485} {\bibfield  {journal} {\bibinfo  {journal}
  {Journal of Fluid Mechanics}\ }\textbf {\bibinfo {volume} {516}},\ \bibinfo
  {pages} {181--209} (\bibinfo {year} {2004})}\BibitemShut {NoStop}%
\bibitem [{\citenamefont {Hughes}\ and\ \citenamefont
  {Griffiths}(2008)}]{hughes2008horizontal}%
  \BibitemOpen
  \bibfield  {author} {\bibinfo {author} {\bibfnamefont {G.~O.}\ \bibnamefont
  {Hughes}}\ and\ \bibinfo {author} {\bibfnamefont {R.~W.}\ \bibnamefont
  {Griffiths}},\ }\bibfield  {title} {\enquote {\bibinfo {title} {Horizontal
  convection},}\ }\href@noop {} {\bibfield  {journal} {\bibinfo  {journal}
  {Annu. Rev. Fluid Mech.}\ }\textbf {\bibinfo {volume} {40}},\ \bibinfo
  {pages} {185--208} (\bibinfo {year} {2008})}\BibitemShut {NoStop}%
\bibitem [{\citenamefont {Shishkina}, \citenamefont {Grossmann},\ and\
  \citenamefont {Lohse}(2016)}]{shishkina2016heat}%
  \BibitemOpen
  \bibfield  {author} {\bibinfo {author} {\bibfnamefont {O.}~\bibnamefont
  {Shishkina}}, \bibinfo {author} {\bibfnamefont {S.}~\bibnamefont
  {Grossmann}}, \ and\ \bibinfo {author} {\bibfnamefont {D.}~\bibnamefont
  {Lohse}},\ }\bibfield  {title} {\enquote {\bibinfo {title} {Heat and momentum
  transport scalings in horizontal convection},}\ }\href@noop {} {\bibfield
  {journal} {\bibinfo  {journal} {Geophysical research letters}\ }\textbf
  {\bibinfo {volume} {43}},\ \bibinfo {pages} {1219--1225} (\bibinfo {year}
  {2016})}\BibitemShut {NoStop}%
\bibitem [{\citenamefont {{Torrance}}, \citenamefont {{Orloff}},\ and\
  \citenamefont {{Rockett}}(1969)}]{Torrance1969}%
  \BibitemOpen
  \bibfield  {author} {\bibinfo {author} {\bibfnamefont {K.~E.}\ \bibnamefont
  {{Torrance}}}, \bibinfo {author} {\bibfnamefont {L.}~\bibnamefont
  {{Orloff}}}, \ and\ \bibinfo {author} {\bibfnamefont {J.~A.}\ \bibnamefont
  {{Rockett}}},\ }\bibfield  {title} {\enquote {\bibinfo {title} {{Experiments
  on natural convection in enclosures with localized heating from below}},}\
  }\href@noop {} {\bibfield  {journal} {\bibinfo  {journal} {Journal of Fluid
  Mechanics}\ }\textbf {\bibinfo {volume} {36}},\ \bibinfo {pages} {21--31}
  (\bibinfo {year} {1969})}\BibitemShut {NoStop}%
\bibitem [{\citenamefont {Torrance}(1979)}]{1979Torrance}%
  \BibitemOpen
  \bibfield  {author} {\bibinfo {author} {\bibfnamefont {K.~E.}\ \bibnamefont
  {Torrance}},\ }\bibfield  {title} {\enquote {\bibinfo {title} {{Natural
  convection in thermally stratified enclosures with localized heating from
  below}},}\ }\href {\doibase 10.1017/S0022112079001567} {\bibfield  {journal}
  {\bibinfo  {journal} {Journal of Fluid Mechanics}\ }\textbf {\bibinfo
  {volume} {95}},\ \bibinfo {pages} {477--495} (\bibinfo {year}
  {1979})}\BibitemShut {NoStop}%
\bibitem [{\citenamefont {Miroshnichenko}\ and\ \citenamefont
  {Sheremet}(2018)}]{miroshnichenko2018turbulent}%
  \BibitemOpen
  \bibfield  {author} {\bibinfo {author} {\bibfnamefont {I.}~\bibnamefont
  {Miroshnichenko}}\ and\ \bibinfo {author} {\bibfnamefont {M.}~\bibnamefont
  {Sheremet}},\ }\bibfield  {title} {\enquote {\bibinfo {title} {Turbulent
  natural convection heat transfer in rectangular enclosures using experimental
  and numerical approaches: A review},}\ }\href@noop {} {\bibfield  {journal}
  {\bibinfo  {journal} {Renewable and Sustainable Energy Reviews}\ }\textbf
  {\bibinfo {volume} {82}},\ \bibinfo {pages} {40--59} (\bibinfo {year}
  {2018})}\BibitemShut {NoStop}%
\bibitem [{\citenamefont {Vasiliev}\ \emph {et~al.}(2016)\citenamefont
  {Vasiliev}, \citenamefont {Sukhanovskii}, \citenamefont {Frick},
  \citenamefont {Budnikov}, \citenamefont {Fomichev}, \citenamefont
  {Bolshukhin},\ and\ \citenamefont {Romanov}}]{vasiliev2016high}%
  \BibitemOpen
  \bibfield  {author} {\bibinfo {author} {\bibfnamefont {A.}~\bibnamefont
  {Vasiliev}}, \bibinfo {author} {\bibfnamefont {A.}~\bibnamefont
  {Sukhanovskii}}, \bibinfo {author} {\bibfnamefont {P.}~\bibnamefont {Frick}},
  \bibinfo {author} {\bibfnamefont {A.}~\bibnamefont {Budnikov}}, \bibinfo
  {author} {\bibfnamefont {V.}~\bibnamefont {Fomichev}}, \bibinfo {author}
  {\bibfnamefont {M.}~\bibnamefont {Bolshukhin}}, \ and\ \bibinfo {author}
  {\bibfnamefont {R.}~\bibnamefont {Romanov}},\ }\bibfield  {title} {\enquote
  {\bibinfo {title} {High rayleigh number convection in a cubic cell with
  adiabatic sidewalls},}\ }\href@noop {} {\bibfield  {journal} {\bibinfo
  {journal} {International Journal of Heat and Mass Transfer}\ }\textbf
  {\bibinfo {volume} {102}},\ \bibinfo {pages} {201--212} (\bibinfo {year}
  {2016})}\BibitemShut {NoStop}%
\bibitem [{\citenamefont {Toppaladoddi}, \citenamefont {Succi},\ and\
  \citenamefont {Wettlaufer}(2017)}]{toppaladoddi2017roughness}%
  \BibitemOpen
  \bibfield  {author} {\bibinfo {author} {\bibfnamefont {S.}~\bibnamefont
  {Toppaladoddi}}, \bibinfo {author} {\bibfnamefont {S.}~\bibnamefont {Succi}},
  \ and\ \bibinfo {author} {\bibfnamefont {J.~S.}\ \bibnamefont {Wettlaufer}},\
  }\bibfield  {title} {\enquote {\bibinfo {title} {Roughness as a route to the
  ultimate regime of thermal convection},}\ }\href@noop {} {\bibfield
  {journal} {\bibinfo  {journal} {Physical review letters}\ }\textbf {\bibinfo
  {volume} {118}},\ \bibinfo {pages} {074503} (\bibinfo {year}
  {2017})}\BibitemShut {NoStop}%
\bibitem [{\citenamefont {Jiang}\ \emph {et~al.}(2019)\citenamefont {Jiang},
  \citenamefont {Zhu}, \citenamefont {Mathai}, \citenamefont {Yang},
  \citenamefont {Verzicco}, \citenamefont {Lohse},\ and\ \citenamefont
  {Sun}}]{jiang2019convective}%
  \BibitemOpen
  \bibfield  {author} {\bibinfo {author} {\bibfnamefont {H.}~\bibnamefont
  {Jiang}}, \bibinfo {author} {\bibfnamefont {X.}~\bibnamefont {Zhu}}, \bibinfo
  {author} {\bibfnamefont {V.}~\bibnamefont {Mathai}}, \bibinfo {author}
  {\bibfnamefont {X.}~\bibnamefont {Yang}}, \bibinfo {author} {\bibfnamefont
  {R.}~\bibnamefont {Verzicco}}, \bibinfo {author} {\bibfnamefont
  {D.}~\bibnamefont {Lohse}}, \ and\ \bibinfo {author} {\bibfnamefont
  {C.}~\bibnamefont {Sun}},\ }\bibfield  {title} {\enquote {\bibinfo {title}
  {Convective heat transfer along ratchet surfaces in vertical natural
  convection},}\ }\href@noop {} {\bibfield  {journal} {\bibinfo  {journal}
  {Journal of fluid mechanics}\ }\textbf {\bibinfo {volume} {873}},\ \bibinfo
  {pages} {1055--1071} (\bibinfo {year} {2019})}\BibitemShut {NoStop}%
\bibitem [{\citenamefont {Dong}\ \emph {et~al.}(2020)\citenamefont {Dong},
  \citenamefont {Wang}, \citenamefont {Dong}, \citenamefont {Huang},
  \citenamefont {Jiang}, \citenamefont {Liu}, \citenamefont {Lu}, \citenamefont
  {Qiu}, \citenamefont {Tang},\ and\ \citenamefont {Zhou}}]{dong2020influence}%
  \BibitemOpen
  \bibfield  {author} {\bibinfo {author} {\bibfnamefont {D.-L.}\ \bibnamefont
  {Dong}}, \bibinfo {author} {\bibfnamefont {B.-F.}\ \bibnamefont {Wang}},
  \bibinfo {author} {\bibfnamefont {Y.-H.}\ \bibnamefont {Dong}}, \bibinfo
  {author} {\bibfnamefont {Y.-X.}\ \bibnamefont {Huang}}, \bibinfo {author}
  {\bibfnamefont {N.}~\bibnamefont {Jiang}}, \bibinfo {author} {\bibfnamefont
  {Y.-L.}\ \bibnamefont {Liu}}, \bibinfo {author} {\bibfnamefont {Z.-M.}\
  \bibnamefont {Lu}}, \bibinfo {author} {\bibfnamefont {X.}~\bibnamefont
  {Qiu}}, \bibinfo {author} {\bibfnamefont {Z.-Q.}\ \bibnamefont {Tang}}, \
  and\ \bibinfo {author} {\bibfnamefont {Q.}~\bibnamefont {Zhou}},\ }\bibfield
  {title} {\enquote {\bibinfo {title} {Influence of spatial arrangements of
  roughness elements on turbulent rayleigh-b{\'e}nard convection},}\
  }\href@noop {} {\bibfield  {journal} {\bibinfo  {journal} {Physics of
  Fluids}\ }\textbf {\bibinfo {volume} {32}},\ \bibinfo {pages} {045114}
  (\bibinfo {year} {2020})}\BibitemShut {NoStop}%
\bibitem [{\citenamefont {Malkus}(1954)}]{malkus1954heat}%
  \BibitemOpen
  \bibfield  {author} {\bibinfo {author} {\bibfnamefont {W.~V.}\ \bibnamefont
  {Malkus}},\ }\bibfield  {title} {\enquote {\bibinfo {title} {The heat
  transport and spectrum of thermal turbulence},}\ }\href@noop {} {\bibfield
  {journal} {\bibinfo  {journal} {Proceedings of the Royal Society of London.
  Series A. Mathematical and Physical Sciences}\ }\textbf {\bibinfo {volume}
  {225}},\ \bibinfo {pages} {196--212} (\bibinfo {year} {1954})}\BibitemShut
  {NoStop}%
\bibitem [{\citenamefont {Evgrafova}\ and\ \citenamefont
  {Sukhanovskii}(2019)}]{evgrafova2019specifics}%
  \BibitemOpen
  \bibfield  {author} {\bibinfo {author} {\bibfnamefont {A.}~\bibnamefont
  {Evgrafova}}\ and\ \bibinfo {author} {\bibfnamefont {A.}~\bibnamefont
  {Sukhanovskii}},\ }\bibfield  {title} {\enquote {\bibinfo {title} {Specifics
  of heat flux from localized heater in a cylindrical layer},}\ }\href@noop {}
  {\bibfield  {journal} {\bibinfo  {journal} {International Journal of Heat and
  Mass Transfer}\ }\textbf {\bibinfo {volume} {135}},\ \bibinfo {pages}
  {761--768} (\bibinfo {year} {2019})}\BibitemShut {NoStop}%
\bibitem [{\citenamefont {Wright}\ \emph {et~al.}(2017)\citenamefont {Wright},
  \citenamefont {Su}, \citenamefont {Scolan}, \citenamefont {Young},\ and\
  \citenamefont {Read}}]{wright2017regimes}%
  \BibitemOpen
  \bibfield  {author} {\bibinfo {author} {\bibfnamefont {S.}~\bibnamefont
  {Wright}}, \bibinfo {author} {\bibfnamefont {S.}~\bibnamefont {Su}}, \bibinfo
  {author} {\bibfnamefont {H.}~\bibnamefont {Scolan}}, \bibinfo {author}
  {\bibfnamefont {R.}~\bibnamefont {Young}}, \ and\ \bibinfo {author}
  {\bibfnamefont {P.~L.}\ \bibnamefont {Read}},\ }\bibfield  {title} {\enquote
  {\bibinfo {title} {Regimes of axisymmetric flow and scaling laws in a
  rotating annulus with local convective forcing},}\ }\href@noop {} {\bibfield
  {journal} {\bibinfo  {journal} {Fluids}\ }\textbf {\bibinfo {volume} {2}},\
  \bibinfo {pages} {41} (\bibinfo {year} {2017})}\BibitemShut {NoStop}%
\bibitem [{\citenamefont {Sukhanovsky}\ \emph {et~al.}(2012)\citenamefont
  {Sukhanovsky}, \citenamefont {Batalov}, \citenamefont {Teymurazov},\ and\
  \citenamefont {Frick}}]{sukhanovsky2012horizontal}%
  \BibitemOpen
  \bibfield  {author} {\bibinfo {author} {\bibfnamefont {A.}~\bibnamefont
  {Sukhanovsky}}, \bibinfo {author} {\bibfnamefont {V.}~\bibnamefont
  {Batalov}}, \bibinfo {author} {\bibfnamefont {A.}~\bibnamefont {Teymurazov}},
  \ and\ \bibinfo {author} {\bibfnamefont {P.}~\bibnamefont {Frick}},\
  }\bibfield  {title} {\enquote {\bibinfo {title} {Horizontal rolls in
  convective flow above a partially heated surface},}\ }\href@noop {}
  {\bibfield  {journal} {\bibinfo  {journal} {The European Physical Journal B}\
  }\textbf {\bibinfo {volume} {85}},\ \bibinfo {pages} {9} (\bibinfo {year}
  {2012})}\BibitemShut {NoStop}%
\bibitem [{\citenamefont {Castaing}\ \emph {et~al.}(1989)\citenamefont
  {Castaing}, \citenamefont {Gunaratne}, \citenamefont {Heslot}, \citenamefont
  {Kadanoff}, \citenamefont {Libchaber}, \citenamefont {Thomae}, \citenamefont
  {Wu}, \citenamefont {Zaleski},\ and\ \citenamefont
  {Zanetti}}]{castaing1989scaling}%
  \BibitemOpen
  \bibfield  {author} {\bibinfo {author} {\bibfnamefont {B.}~\bibnamefont
  {Castaing}}, \bibinfo {author} {\bibfnamefont {G.}~\bibnamefont {Gunaratne}},
  \bibinfo {author} {\bibfnamefont {F.}~\bibnamefont {Heslot}}, \bibinfo
  {author} {\bibfnamefont {L.}~\bibnamefont {Kadanoff}}, \bibinfo {author}
  {\bibfnamefont {A.}~\bibnamefont {Libchaber}}, \bibinfo {author}
  {\bibfnamefont {S.}~\bibnamefont {Thomae}}, \bibinfo {author} {\bibfnamefont
  {X.-Z.}\ \bibnamefont {Wu}}, \bibinfo {author} {\bibfnamefont
  {S.}~\bibnamefont {Zaleski}}, \ and\ \bibinfo {author} {\bibfnamefont
  {G.}~\bibnamefont {Zanetti}},\ }\bibfield  {title} {\enquote {\bibinfo
  {title} {Scaling of hard thermal turbulence in rayleigh-b{\'e}nard
  convection},}\ }\href@noop {} {\bibfield  {journal} {\bibinfo  {journal}
  {Journal of Fluid Mechanics}\ }\textbf {\bibinfo {volume} {204}},\ \bibinfo
  {pages} {1--30} (\bibinfo {year} {1989})}\BibitemShut {NoStop}%
\bibitem [{\citenamefont {Chavanne}\ \emph {et~al.}(2001)\citenamefont
  {Chavanne}, \citenamefont {Chilla}, \citenamefont {Chabaud}, \citenamefont
  {Castaing},\ and\ \citenamefont {Hebral}}]{chavanne2001turbulent}%
  \BibitemOpen
  \bibfield  {author} {\bibinfo {author} {\bibfnamefont {X.}~\bibnamefont
  {Chavanne}}, \bibinfo {author} {\bibfnamefont {F.}~\bibnamefont {Chilla}},
  \bibinfo {author} {\bibfnamefont {B.}~\bibnamefont {Chabaud}}, \bibinfo
  {author} {\bibfnamefont {B.}~\bibnamefont {Castaing}}, \ and\ \bibinfo
  {author} {\bibfnamefont {B.}~\bibnamefont {Hebral}},\ }\bibfield  {title}
  {\enquote {\bibinfo {title} {Turbulent rayleigh--b{\'e}nard convection in
  gaseous and liquid he},}\ }\href@noop {} {\bibfield  {journal} {\bibinfo
  {journal} {Physics of Fluids}\ }\textbf {\bibinfo {volume} {13}},\ \bibinfo
  {pages} {1300--1320} (\bibinfo {year} {2001})}\BibitemShut {NoStop}%
\bibitem [{\citenamefont {Golitsyn}(1979)}]{golitsyn1979simple}%
  \BibitemOpen
  \bibfield  {author} {\bibinfo {author} {\bibfnamefont {G.}~\bibnamefont
  {Golitsyn}},\ }\bibfield  {title} {\enquote {\bibinfo {title} {Simple
  theoretical and experimental study of convection with some geophysical
  applications and analogies},}\ }\href@noop {} {\bibfield  {journal} {\bibinfo
   {journal} {Journal of Fluid Mechanics}\ }\textbf {\bibinfo {volume} {95}},\
  \bibinfo {pages} {567--608} (\bibinfo {year} {1979})}\BibitemShut {NoStop}%
\bibitem [{\citenamefont {Ripesi}\ \emph {et~al.}(2014)\citenamefont {Ripesi},
  \citenamefont {Biferale}, \citenamefont {Sbragaglia},\ and\ \citenamefont
  {Wirth}}]{ripesi2014natural}%
  \BibitemOpen
  \bibfield  {author} {\bibinfo {author} {\bibfnamefont {P.}~\bibnamefont
  {Ripesi}}, \bibinfo {author} {\bibfnamefont {L.}~\bibnamefont {Biferale}},
  \bibinfo {author} {\bibfnamefont {M.}~\bibnamefont {Sbragaglia}}, \ and\
  \bibinfo {author} {\bibfnamefont {A.}~\bibnamefont {Wirth}},\ }\bibfield
  {title} {\enquote {\bibinfo {title} {Natural convection with mixed insulating
  and conducting boundary conditions: low-and high-rayleigh-number regimes},}\
  }\href@noop {} {\bibfield  {journal} {\bibinfo  {journal} {Journal of Fluid
  Mechanics}\ }\textbf {\bibinfo {volume} {742}},\ \bibinfo {pages} {636--663}
  (\bibinfo {year} {2014})}\BibitemShut {NoStop}%
\bibitem [{\citenamefont {Wang}, \citenamefont {Huang},\ and\ \citenamefont
  {Xia}(2017)}]{wang2017thermal}%
  \BibitemOpen
  \bibfield  {author} {\bibinfo {author} {\bibfnamefont {F.}~\bibnamefont
  {Wang}}, \bibinfo {author} {\bibfnamefont {S.-D.}\ \bibnamefont {Huang}}, \
  and\ \bibinfo {author} {\bibfnamefont {K.-Q.}\ \bibnamefont {Xia}},\
  }\bibfield  {title} {\enquote {\bibinfo {title} {Thermal convection with
  mixed thermal boundary conditions: effects of insulating lids at the top},}\
  }\href@noop {} {\bibfield  {journal} {\bibinfo  {journal} {Journal of Fluid
  Mechanics}\ }\textbf {\bibinfo {volume} {817}} (\bibinfo {year}
  {2017})}\BibitemShut {NoStop}%
\bibitem [{\citenamefont {Bakhuis}\ \emph {et~al.}(2018)\citenamefont
  {Bakhuis}, \citenamefont {Ostilla-M{\'o}nico}, \citenamefont {Van Der~Poel},
  \citenamefont {Verzicco},\ and\ \citenamefont {Lohse}}]{bakhuis2018mixed}%
  \BibitemOpen
  \bibfield  {author} {\bibinfo {author} {\bibfnamefont {D.}~\bibnamefont
  {Bakhuis}}, \bibinfo {author} {\bibfnamefont {R.}~\bibnamefont
  {Ostilla-M{\'o}nico}}, \bibinfo {author} {\bibfnamefont {E.~P.}\ \bibnamefont
  {Van Der~Poel}}, \bibinfo {author} {\bibfnamefont {R.}~\bibnamefont
  {Verzicco}}, \ and\ \bibinfo {author} {\bibfnamefont {D.}~\bibnamefont
  {Lohse}},\ }\bibfield  {title} {\enquote {\bibinfo {title} {Mixed insulating
  and conducting thermal boundary conditions in rayleigh--b{\'e}nard
  convection},}\ }\href@noop {} {\bibfield  {journal} {\bibinfo  {journal}
  {Journal of fluid mechanics}\ }\textbf {\bibinfo {volume} {835}},\ \bibinfo
  {pages} {491--511} (\bibinfo {year} {2018})}\BibitemShut {NoStop}%
\bibitem [{\citenamefont {Batalov}, \citenamefont {Sukhanovsky},\ and\
  \citenamefont {Frick}(2010)}]{batalov2010laboratory}%
  \BibitemOpen
  \bibfield  {author} {\bibinfo {author} {\bibfnamefont {V.}~\bibnamefont
  {Batalov}}, \bibinfo {author} {\bibfnamefont {A.}~\bibnamefont
  {Sukhanovsky}}, \ and\ \bibinfo {author} {\bibfnamefont {P.}~\bibnamefont
  {Frick}},\ }\bibfield  {title} {\enquote {\bibinfo {title} {Laboratory study
  of differential rotation in a convective rotating layer},}\ }\href@noop {}
  {\bibfield  {journal} {\bibinfo  {journal} {Geophysical and Astrophysical
  Fluid Dynamics}\ }\textbf {\bibinfo {volume} {104}},\ \bibinfo {pages}
  {349--368} (\bibinfo {year} {2010})}\BibitemShut {NoStop}%
\bibitem [{\citenamefont {Sukhanovskii}, \citenamefont {Evgrafova},\ and\
  \citenamefont {Popova}(2016{\natexlab{a}})}]{sukhanovskii2016}%
  \BibitemOpen
  \bibfield  {author} {\bibinfo {author} {\bibfnamefont {A.}~\bibnamefont
  {Sukhanovskii}}, \bibinfo {author} {\bibfnamefont {A.}~\bibnamefont
  {Evgrafova}}, \ and\ \bibinfo {author} {\bibfnamefont {E.}~\bibnamefont
  {Popova}},\ }\bibfield  {title} {\enquote {\bibinfo {title} {Horizontal rolls
  over localized heat source in a cylindrical layer},}\ }\href@noop {}
  {\bibfield  {journal} {\bibinfo  {journal} {Physica D: Nonlinear Phenomena}\
  }\textbf {\bibinfo {volume} {316}},\ \bibinfo {pages} {23--33} (\bibinfo
  {year} {2016}{\natexlab{a}})}\BibitemShut {NoStop}%
\bibitem [{\citenamefont {Sukhanovskii}, \citenamefont {Evgrafova},\ and\
  \citenamefont {Popova}(2016{\natexlab{b}})}]{sukhanovskii2016laboratory}%
  \BibitemOpen
  \bibfield  {author} {\bibinfo {author} {\bibfnamefont {A.}~\bibnamefont
  {Sukhanovskii}}, \bibinfo {author} {\bibfnamefont {A.}~\bibnamefont
  {Evgrafova}}, \ and\ \bibinfo {author} {\bibfnamefont {E.}~\bibnamefont
  {Popova}},\ }\bibfield  {title} {\enquote {\bibinfo {title} {Laboratory study
  of a steady-state convective cyclonic vortex},}\ }\href@noop {} {\bibfield
  {journal} {\bibinfo  {journal} {Quarterly Journal of the Royal Meteorological
  Society}\ }\textbf {\bibinfo {volume} {142}},\ \bibinfo {pages} {2214--2223}
  (\bibinfo {year} {2016}{\natexlab{b}})}\BibitemShut {NoStop}%
\bibitem [{\citenamefont {Sukhanovskii}, \citenamefont {Evgrafova},\ and\
  \citenamefont {Popova}(2017)}]{sukhanovskii2017non}%
  \BibitemOpen
  \bibfield  {author} {\bibinfo {author} {\bibfnamefont {A.}~\bibnamefont
  {Sukhanovskii}}, \bibinfo {author} {\bibfnamefont {A.}~\bibnamefont
  {Evgrafova}}, \ and\ \bibinfo {author} {\bibfnamefont {E.}~\bibnamefont
  {Popova}},\ }\bibfield  {title} {\enquote {\bibinfo {title} {Non-axisymmetric
  structure of the boundary layer of intensive cyclonic vortex},}\ }\href@noop
  {} {\bibfield  {journal} {\bibinfo  {journal} {Dynamics of Atmospheres and
  Oceans}\ }\textbf {\bibinfo {volume} {80}},\ \bibinfo {pages} {12--28}
  (\bibinfo {year} {2017})}\BibitemShut {NoStop}%
\bibitem [{\citenamefont {Sukhanovskii}\ and\ \citenamefont
  {Popova}(2020)}]{sukhanovskii2020importance}%
  \BibitemOpen
  \bibfield  {author} {\bibinfo {author} {\bibfnamefont {A.}~\bibnamefont
  {Sukhanovskii}}\ and\ \bibinfo {author} {\bibfnamefont {E.}~\bibnamefont
  {Popova}},\ }\bibfield  {title} {\enquote {\bibinfo {title} {The importance
  of horizontal rolls in the rapid intensification of tropical cyclones},}\
  }\href@noop {} {\bibfield  {journal} {\bibinfo  {journal} {Boundary-Layer
  Meteorology}\ ,\ \bibinfo {pages} {1--18}} (\bibinfo {year}
  {2020})}\BibitemShut {NoStop}%
\bibitem [{\citenamefont {Batalov}, \citenamefont {Sukhanovskii},\ and\
  \citenamefont {Frik}(2007)}]{batalov2007experimental}%
  \BibitemOpen
  \bibfield  {author} {\bibinfo {author} {\bibfnamefont {V.}~\bibnamefont
  {Batalov}}, \bibinfo {author} {\bibfnamefont {A.}~\bibnamefont
  {Sukhanovskii}}, \ and\ \bibinfo {author} {\bibfnamefont {P.}~\bibnamefont
  {Frik}},\ }\bibfield  {title} {\enquote {\bibinfo {title} {Experimental
  investigation of helicoidal rolls in an advective flow over a hot horizontal
  surface},}\ }\href@noop {} {\bibfield  {journal} {\bibinfo  {journal} {Fluid
  Dynamics}\ }\textbf {\bibinfo {volume} {42}},\ \bibinfo {pages} {540--549}
  (\bibinfo {year} {2007})}\BibitemShut {NoStop}%
\bibitem [{\citenamefont {Waleffe}, \citenamefont {Boonkasame},\ and\
  \citenamefont {Smith}(2015)}]{waleffe2015heat}%
  \BibitemOpen
  \bibfield  {author} {\bibinfo {author} {\bibfnamefont {F.}~\bibnamefont
  {Waleffe}}, \bibinfo {author} {\bibfnamefont {A.}~\bibnamefont {Boonkasame}},
  \ and\ \bibinfo {author} {\bibfnamefont {L.~M.}\ \bibnamefont {Smith}},\
  }\bibfield  {title} {\enquote {\bibinfo {title} {Heat transport by coherent
  rayleigh-b{\'e}nard convection},}\ }\href@noop {} {\bibfield  {journal}
  {\bibinfo  {journal} {Physics of Fluids}\ }\textbf {\bibinfo {volume} {27}},\
  \bibinfo {pages} {051702} (\bibinfo {year} {2015})}\BibitemShut {NoStop}%
\bibitem [{\citenamefont {Niemela}\ \emph {et~al.}(2000)\citenamefont
  {Niemela}, \citenamefont {Skrbek}, \citenamefont {Sreenivasan},\ and\
  \citenamefont {Donnelly}}]{niemela2000turbulent}%
  \BibitemOpen
  \bibfield  {author} {\bibinfo {author} {\bibfnamefont {J.}~\bibnamefont
  {Niemela}}, \bibinfo {author} {\bibfnamefont {L.}~\bibnamefont {Skrbek}},
  \bibinfo {author} {\bibfnamefont {K.}~\bibnamefont {Sreenivasan}}, \ and\
  \bibinfo {author} {\bibfnamefont {R.}~\bibnamefont {Donnelly}},\ }\bibfield
  {title} {\enquote {\bibinfo {title} {Turbulent convection at very high
  rayleigh numbers},}\ }\href@noop {} {\bibfield  {journal} {\bibinfo
  {journal} {Nature}\ }\textbf {\bibinfo {volume} {404}},\ \bibinfo {pages}
  {837--840} (\bibinfo {year} {2000})}\BibitemShut {NoStop}%
\bibitem [{\citenamefont {He}\ \emph {et~al.}(2012)\citenamefont {He},
  \citenamefont {Funfschilling}, \citenamefont {Nobach}, \citenamefont
  {Bodenschatz},\ and\ \citenamefont {Ahlers}}]{he2012transition}%
  \BibitemOpen
  \bibfield  {author} {\bibinfo {author} {\bibfnamefont {X.}~\bibnamefont
  {He}}, \bibinfo {author} {\bibfnamefont {D.}~\bibnamefont {Funfschilling}},
  \bibinfo {author} {\bibfnamefont {H.}~\bibnamefont {Nobach}}, \bibinfo
  {author} {\bibfnamefont {E.}~\bibnamefont {Bodenschatz}}, \ and\ \bibinfo
  {author} {\bibfnamefont {G.}~\bibnamefont {Ahlers}},\ }\bibfield  {title}
  {\enquote {\bibinfo {title} {Transition to the ultimate state of turbulent
  rayleigh-b{\'e}nard convection},}\ }\href@noop {} {\bibfield  {journal}
  {\bibinfo  {journal} {Physical review letters}\ }\textbf {\bibinfo {volume}
  {108}},\ \bibinfo {pages} {024502} (\bibinfo {year} {2012})}\BibitemShut
  {NoStop}%
\bibitem [{\citenamefont {Cioni}, \citenamefont {Ciliberto},\ and\
  \citenamefont {Sommeria}(1997)}]{cioni1997strongly}%
  \BibitemOpen
  \bibfield  {author} {\bibinfo {author} {\bibfnamefont {S.}~\bibnamefont
  {Cioni}}, \bibinfo {author} {\bibfnamefont {S.}~\bibnamefont {Ciliberto}}, \
  and\ \bibinfo {author} {\bibfnamefont {J.}~\bibnamefont {Sommeria}},\
  }\bibfield  {title} {\enquote {\bibinfo {title} {Strongly turbulent
  rayleigh--b{\'e}nard convection in mercury: comparison with results at
  moderate prandtl number},}\ }\href@noop {} {\bibfield  {journal} {\bibinfo
  {journal} {Journal of Fluid Mechanics}\ }\textbf {\bibinfo {volume} {335}},\
  \bibinfo {pages} {111--140} (\bibinfo {year} {1997})}\BibitemShut {NoStop}%
\end{thebibliography}%

\end{document}